\documentclass[sigconf]{acmart}
\usepackage{natbib}
\usepackage{graphicx}
\usepackage{textcomp}
\usepackage{float}
\usepackage[hang,flushmargin]{footmisc}
\usepackage{import}
\usepackage{color}

\usepackage{soul}
\usepackage{pifont}
\usepackage{enumitem}
\usepackage{multirow}
\usepackage{blindtext}
\usepackage{dirtytalk} 

\usepackage{tabularx} 
\usepackage{multirow}
\usepackage{listings} 
\usepackage{todonotes}
\lstdefinestyle{python}{ 
    xleftmargin=6.0ex,
    xrightmargin=2.0ex,
    numbers=left,
    frame=single
}
\definecolor{gray10}{gray}{.9}
\definecolor{arsenic}{rgb}{0.23, 0.27, 0.29}
\usepackage{color}

\RequirePackage{fontawesome}

\definecolor{gray50}{gray}{.5}
\definecolor{gray40}{gray}{.6}
\definecolor{gray30}{gray}{.7}
\definecolor{gray20}{gray}{.8}
\definecolor{gray10}{gray}{.9}
\definecolor{gray05}{gray}{.95}

\newlength\Linewidth
\def\findlength{\setlength\Linewidth\linewidth
  \addtolength\Linewidth{-4\fboxrule}
  \addtolength\Linewidth{-3\fboxsep}
}
\newenvironment{examplebox}{\par\begingroup
  \setlength{\fboxsep}{5pt}\findlength
  \setbox0=\vbox\bgroup\noindent
  \hsize=0.95\linewidth
  \begin{minipage}{0.95\linewidth}\normalsize}
  {\end{minipage}\egroup
  \textcolor{gray20}{\fboxsep1.5pt\fbox
    {\fboxsep5pt\colorbox{gray05}{\normalcolor\box0}}}
  \endgroup\par\noindent
  \normalcolor\ignorespacesafterend}

\begin{document}

\title{Towards a Technical Debt for AI-based Recommender System}

\author{Sergio Moreschini}
\affiliation{%
  \institution{Tampere University}
  \city{Tampere}
  \country{Finland}
}
\email{sergio.moreschini@tuni.fi}

\author{Ludovik Coba}
\authornote{This work was completed before joining Roku Inc.}
\affiliation{%
  \institution{Roku Inc.}
  \city{London}
  \country{UK}
}
\email{lcoba@roku.com}

\author{Valentina Lenarduzzi}
\affiliation{%
  \institution{University of Oulu}
  \city{Oulu}
  \country{Finland}
}
\email{valentina.lenarduzzi@oulu.fi}

\begin{abstract}

Balancing the management of technical debt within recommender systems requires effectively juggling the introduction of new features with the ongoing maintenance and enhancement of the current system. Within the realm of recommender systems, technical debt encompasses the trade-offs and expedient choices made during the development and upkeep of the recommendation system, which could potentially have adverse effects on its long-term performance, scalability, and maintainability. In this vision paper, our objective is to kickstart a research direction regarding Technical Debt in AI-based Recommender Systems. We identified 15 potential factors, along with detailed explanations outlining why it is advisable to consider them.
\end{abstract}




\maketitle

\section{Introduction}
\label{sec:Introduction}

Managing technical debt in recommender systems involves a balance between delivering new features and maintaining and improving the existing system~\cite{Cunningham1992, DagstuhlReport}. Regularly reviewing and addressing these debt areas is essential to ensure that the AI-based Recommender Systems continues to provide high-quality recommendations and remains adaptable to changing user needs and market conditions~\cite{LenarduzziSLR2021}. 

Technical Debt in the context of AI-based Recommender Systems refers to the shortcuts or compromises made during the development and maintenance of the recommendation system that may negatively impact its long-term performance, scalability, and maintainability. Like in any software development project, Technical Debt in AI-based Recommender Systems can accumulate over time and should be managed carefully to avoid future issues.

The accumulation of Technical Debt in AI-based Recommender Systems can result in degraded user experiences, increased development and maintenance costs, and difficulties in remaining competitive in the fast-evolving field of recommendation technology. Therefore, it's crucial for organizations to prioritize debt reduction and invest in the long-term health of their Recommender Systems.

In this ongoing research papers, we aim at initiating the discussion about a specific Technical Debt for AI-based Recommender Systems. We pointed 15 potential factors and we provide explanations for why we recommend taking them into account. 
Moreover, we define a roadmap with the future steps we want to investigate. 

\textbf{Paper Structure}.  Section \ref{sec:TD} introduces the concept of Technical Debt while Section \ref{sec:RS} the recommender systems. Section \ref{sec:Idea} describes our research vision idea, Section \ref{sec:Roadmap} presents our roadmap.  Section \ref{sec:Conclusion} draws some conclusions.

\section{Technical Debt}
\label{sec:TD}

Technical debt is a concept in software engineering that refers to the implied cost of additional work that arises when software is developed or maintained quickly and with shortcuts or suboptimal solutions. It's a metaphorical way of describing the trade-off between getting a piece of software out quickly and dealing with the long-term consequences of taking shortcuts or making compromises in the development process~\cite{Cunningham1992, DagstuhlReport}.

Technical debt is a concept that highlights the consequences of taking shortcuts or making compromises during software development. It emphasizes the importance of managing and reducing debt over time to ensure the long-term quality and maintainability of software systems~\cite{LenarduzziSLR2021}. 

Unfortunately, Technical Debt is an unavoidable but beneficial aspect under certain circumstances, often linked to unforeseeable internal or external business or environmental forces~\cite{MARTINI2015237}. 

Evidence about the good and bad effects has been investigated in the last 10 years. Researchers proposed initial approaches to manage and reduce Technical Debt, developing also some tools, but more research and insights are needed~\cite{Ciolkowski2021}.
Technical Debt is composed of ``technical issues'' that include any kind of information derived from the source code and from the software process, such as usage of specific patterns, compliance with coding or documentation conventions, and architectural issues~\cite{Lenarduzzi2019}. 

However, just like any other financial debt, every Technical Debt has an interest attached, or else an extra cost or negative impact that is generated by the presence of a sub-optimal solution~\cite{Li2015}. When such interest becomes very costly, it can lead to disruptive events, such as development crises~\cite{MARTINI2015237}. The current best practices employed by software companies include keeping Technical Debt at bay by avoiding it if the consequences are known or refactoring or rewriting code and other artifacts in order to get rid of the accumulated sub-optimal solutions and their negative impact. 
Companies cannot afford to avoid or repay all the Technical Debt that is generated continuously and may be unknown~\cite{MARTINI2015237}. The main business goals of companies are to continuously deliver value to their customers and to maintain their products.

\section{Recommeder Systems}
\label{sec:RS}

RS have become an integral part of modern online platforms and applications, leveraging machine learning techniques to assist users in finding relevant items or content based on their preferences, past behaviors, or similarities to other users. With the vast amount of available options, these systems play a crucial role in enhancing user experience, increasing engagement, and facilitating decision-making \cite{ricci2015recommender, jannach2010recommender}.
RS can be found across a wide range of online platforms and applications, including e-commerce websites, streaming services, social media platforms, news portals, and more. Their primary goal is to help users navigate through an overwhelming sea of choices and discover new products, movies, music, articles, and other items of interest.
Successfully deploying RS in a production environment is not without its challenges. Organizations face various hurdles that need to be overcome to deliver effective and scalable recommendations to users. Some of these challenges include harmonizing multiple data sources, scalability, evaluation, feedback loop, ethical considerations, maintenance and adaptability. Thus, building such large systems could lead to the accumulation of technical debt, which can arise from hurriedly implemented algorithms, sub-optimal data processing methods, or insufficient testing. Over time, this debt can lead to issues such as decreased system performance, increased maintenance efforts, and difficulty in implementing new features or algorithms. Addressing tech debt in RS is crucial to maintaining their effectiveness and ensuring long-term sustainability.

\section{Towards a Technical Debt for AI-based Recommender Systems}
\label{sec:Idea}

In this section, we share 15 possible Technical Debt factors that we discovered in the literature regarding AI-based Recommender Systems. The identification of these factors is the result of our work in collaboration with practitioners in the field.
A rationale for considering each factor is provided.
In order to conceptualize a Technical Debt for a AI-based Recommender System (RS) we need first to identify which can be the factors that could characterize this type of Technical Debt (TD).

\begin{itemize} 
    \item \textbf{Data Contracts:} 
    Data contracts are very important for defining API data models and data governance \cite{lu2022towards}.
    They play a crucial role in ensuring clarity and consistency in data exchange processes between different parties or systems. By defining the structure, format, and exchange rules, they help prevent misunderstandings and ensure seamless communication in distributed data architectures. Particular types of recommenders (e.g. context-aware RS \cite{raza2019progress}) generate live predictions using data that is extremely volatile (e.g. user features), thus, in such a contract, the client and the service provider define the signals (or features) that the model needs to function properly. 
    Often, data contracts are defined before or during implementation. This may lead to a redefinition of the data pipeline or the signals produced at a later stage, resulting in significant time loss in implementing the hydration service to serve missing features or modeling the RS.

    \item \textbf{Glue code:} There is a large plethora of generic packages for RS \cite{ekstrand2011rethinking, pasumarthi2019tf, bayer2016fastfm, coba2022recoxplainer, argyriou2020microsoft}. Using generic packages frequently leads to the creation of a glue code system design pattern. This results in the development of a substantial amount of auxiliary code to facilitate data exchange with these packages. Glue code can be costly in the long term because it tends to lock a system into function only with a specific package, making it expensive to explore alternative solutions. Furthermore, it can hinder improvements, as it becomes challenging to leverage domain-specific characteristics or fine-tune the system to meet specific domain goals.

    \item \textbf{Pipeline Jungles:} ML-based RSs, particularly those following MLOps guidelines, are composed of multiple pipelines which provide Continuous Training, Continuous Deployment, and Continuous Integration. This also requires automation of the data scraping, data management, data preprocessing, and monitoring steps, which results in the creation of additional pipelines. This high amount of pipelines can turn into a complex \textit{jungle} of scraping, joining, and sampling procedures. The result is a domino effect caused by any small change.

     \item \textbf{Dead Experimental Codepaths:} As a consequence of the previous two points, in the short term it becomes more convenient to perform code experimentation by introducing experimental codepaths as conditional branches within the primary production code. The cost of experimentation is relatively low from a TD perspective however, the overall cost will increase due to the accumulation of these unused experimental codepaths. Among these unused codepaths only few in the end will be used while most of them are usually tested and abandoned. The main result of this accumulated cost is presented when trying to recover an abandoned branch which might presents compatibility issues due to unfulfilled requirements.

    \item \textbf{Prototype Smell:} Following the previous point, it is a standard procedure to use branches to create new small-scale environment based on prototypes to test new ideas. Maintaining a prototype environment creates new costs to add to the TD, but the main drawback is the feeling that a prototype, and its environment, are ready for production. 
    In some cases, the new environment is not completely compatible with the main branch creating difficulties when deciding to switch between the main branch and a prototype.

    \item \textbf{Inconsistency between offline and online evaluation results:} Most RS are shipped to production using standard evaluation metrics (such as normalized Discounted Cumulative Gain, recall, precision, etc.) and sometimes, ad-hoc KPI-related metrics \cite{shani2011evaluating}. Using advanced user models we can execute simulations of users interactions with the system, thus reducing the need for expensive user studies and online testing.  However, the approaches are based on simple evaluation proxies that don't map the intricate behaviour of a user, often having offline experimental results not well correlated to online results \cite{argyriou2020microsoft, shani2011evaluating, garcin2014offline, krauth2020offline}.

    \item \textbf{Multiple-Language Smell:} Nowadays, a system can be composed of a combination of several programming languages. Analyzing and maintaining multi-language systems pose significant challenges, and developers consistently encounter difficulties when working with these complex systems~\cite{MultiLanguages}. 
    Some of the consequences of this behaviour are code smells within the source code reflecting suboptimal coding choices that manifest as indicators of the existence of code anti-patterns~\cite{MultiLanguages2}.

     \item \textbf{Conflicting KPI goals:} When defying project goals, a well-known approach involves defining precise Key Performance Indicators (KPIs) that align with these objectives. These KPIs act as quantifiable metrics, enabling the developer to monitor progress and ensure that your endeavours effectively advance the developer toward the profit-related targets. Nevertheless, in certain instances, these KPIs may be mutually contradictory, making it impossible to maximize both simultaneously. For instance, on a search engine, engagement results in long-term decreased conversion and vice-versa. However, more sophisticated metrics could interact in non-obvious ways that are invisible before running online experimentation.

     \cite{shani2011evaluating, abdollahpouri2020multistakeholder}.

    \item \textbf{Plain-Old-Data Type Smell:} RSs receive and provide valuable information, but this information is often represented using basic data types such as raw floats and integers. To ensure the correct and meaningful use of both model parameters and predictions, a well-designed an RS based on ML should not only produce results, but also provide a clear and informative context. This context is essential to make ML systems transparent, interpretable, and reliable to use in practice \cite{hidden}.

    \item \textbf{Reproducibility Debt:} Reproducibility is a hard constraint in software development, but also it has been an emergin topic in RS \cite{ferrari2019we, ccoba2017replication}. When developing RS software it is important to follow specific guidelines to ensure that this and other constraints such as scalability, explainability and interpretability. This is one of the main reasons why MLOps has been emerging nowadays. Compared to classic software development, RS software does not rely only on code but also on data therefore, MLOps has data versioning as one of its foundations to allow reproducibility by reducing to minimum the influence of external factors.

    \item \textbf{Cultural Debt:} Sometimes there is a clear boundary between RS research and engineering, but maintaining this rigid distinction can undermine the long-term health of a system. The cultural debt refers to the debt that is created when different figures (or cultures) cooperate by defining lines and responsibilities. As DevOps has suggested for the last 30 years this is usually not the best approach as attributes should be equally valued and incentivized and not overshadowed by the pursuit of improved results alone. In RS software, it is imperative to cultivate team cultures that not only celebrate advancements in model accuracy but also emphasize the importance of eliminating unnecessary features, reducing complexity, enhancing reproducibility, ensuring system stability, and bolstering monitoring capabilities. As RS are user-feature systems and require the engagement of several teams the cultural debt is therefore more emphasized. 
    
    \item \textbf{Performance Deterioration:} One of the main reasons for performance deterioration are \textit{concept} and \textit{data drift} \cite{gama2014survey, liu2022monolith}. Concept drift arises when the initially established relationship between the target and independent variables, on which a model was initially trained, either becomes obsolete or transforms into an unfamiliar configuration for the model. Concept drift leads to a decline in model accuracy because, during production, the model is unaware of these alterations.
    Data drift refers to the evolving or abrupt changes in the data distribution used to train a machine learning model compared to the data distribution encountered in the actual deployment context. Such changes can be due to a variety of factors, including changes in user actions, changes in the underlying data generation processes, or external events that affect the data collection process.
    Following the previous example for seasonality, if we would train the RS using a dataset based on a moving window, we would see that when training the model including the data for the holiday season the accuracy of the model would decrease due to data drift.
    One of the main factors influencing concept drift for RS is seasonality. An example can be related to how during the holiday season, people are looking for seasonal music, flights, or items in a variety of online services such as music streaming, travel companies, or stores.

    \item \textbf{Feedback Loops:} direct feedback loops have been received a lot of interest in research the recent years \cite{cchaney2018algorithmic,mansoury2020feedback,jiang2019degenerate}. They arise when the results generated by an RS model start to impact the subsequent input features that the same model uses for training. This could result in algorithmic bias (e.g. popularity bias) which intensifies over time as models, because, in many cases, users will be limited to interact with recommendations, which themselves are later used as part of the historical interaction of users, reinforcing the belief of the model about the validity of these recommendation. 
    
    %

    \item \textbf{Hidden Feedback Loops:} hidden feedback loops are harder to establish and consequentially haven't received a lot of research focus in the RS community. Nowadays companies are using multiple RS in parallel and all models could influence the historical interactions that other models will use in training. For instance, Netflix is personalising the artwork presented to users\cite{amat2018artwork}, but also the rows and their content\cite{steck2021deep}, practice which has become an industry standard.
    

\end{itemize}

\section{Roadmap}
\label{sec:Roadmap}
Our roadmap includes the following points: 
\begin{itemize} [leftmargin=*]
    \item \textbf{AI-based Recommender Systems TD factors validation.} We will validate these factors with TD and Recommender systems experts, to determine if: 
\begin{itemize}
     \item They consider these factors a real technical issue (and why)
     \item How severe they consider them
     \item How to fix them (how and when)
\end{itemize}
    \item \textbf{AI-based Recommender Systems TD factors mining.} We will define a method to collect the factors validated by the experts and we collect them in real projects (open source and industrial ones). 
    \item \textbf{AI-based Recommender Systems TD factors impact} We will investigate the impact of the identified factors on: 
    \begin{itemize}
        \item \textit{Software Quality}. We already identified some software quality attributes that can be more negatively impacted by these factors, such as maintainability, readability, and testability. 
        \item \textit{Development Process}. We will measure some software metrics related to the development process such as process metrics proposed by Kamei~\cite{kamei2012large}. 
    \end{itemize}
\end{itemize}

\section{Conclusion}
\label{sec:Conclusion}
In this paper, we raised up the conceptualization of Technical Debt in AI-based Recommender Systems. We have identified 15 potential factors and offered explanations for why we advocate considering them. Additionally, we outlined a roadmap outlining the future steps we intend to explore.

\balance
\bibliographystyle{ACM-Reference-Format}
\bibliography{bibliography.bib}

\end{document}